\documentclass{article}
\usepackage{arxiv}
\usepackage{graphicx}
\usepackage{algorithm}
\usepackage{cite}
\usepackage{balance}
\usepackage{xcolor}
\usepackage{textcomp}
\usepackage{amsmath,amssymb,amsfonts}
\usepackage{algorithmic}
\usepackage[utf8]{inputenc} 
\usepackage[T1]{fontenc}    
\usepackage{hyperref}       
\usepackage{url}            
\usepackage{booktabs}       
\usepackage{amsfonts}       
\usepackage{nicefrac}       
\usepackage{microtype}      
\usepackage{lipsum}		
\usepackage{doi}
\usepackage{pdfpages}
\usepackage{caption}
\usepackage{tabularx} 
\usepackage[thinlines]{easytable}
\usepackage[algo2e]{algorithm2e}
\usepackage{dblfloatfix}
\usepackage{pifont}
\usepackage{placeins}

\title{hChain 4.0: A Secure and Scalable Permissioned Blockchain for EHR Management in Smart Healthcare}

\author{ \href{https://orcid.org/0000-0002-3047-486X}{\includegraphics[scale=0.06]{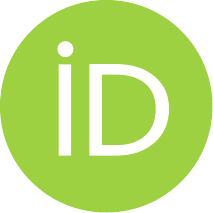}\hspace{1mm}Musharraf N.~Alruwaill} \\
	Department of Computer Science and Engineering \\
	University of North Texas\\
	Denton, TX 76203 \\
	\texttt{MusharrafAlruwaill@my.unt.edu} \\
	\And
	\href{https://orcid.org/0000-0003-2959-6541}{\includegraphics[scale=0.06]{orcid.pdf}\hspace{1mm}Saraju P.~Mohanty} \\
	Department of Computer Science and Engineering \\
	University of North Texas\\
	Denton, TX 76203 \\
	\texttt{saraju.mohanty@unt.edu} \\
	\And
	\href{https://orcid.org/0000-0002-1616-7628}{\includegraphics[scale=0.06]{orcid.pdf}\hspace{1mm}Elias Kougianos} \\
	Department of Electrical Engineering \\
	University of North Texas\\
	Denton, TX 76203 \\
	\texttt{elias.kougianos@unt.edu} \\
}

\renewcommand{\shorttitle}{\textit hChain 4.0: Secure and Scalable Permissioned Blockchain for EHR Management}

\hypersetup{
	pdftitle={A template for the arxiv style},
	pdfsubject={q-bio.NC, q-bio.QM},
	pdfauthor={David S.~Hippocampus, Elias D.~Striatum},
	pdfkeywords={First keyword, Second keyword, More},
}

\begin{document}
	\maketitle
	
	\begin{abstract}
		The growing utilization of Internet of Medical Things (IoMT) devices, including smartwatches and wearable medical devices, has facilitated real-time health monitoring and data analysis to enhance healthcare outcomes. These gadgets necessitate improved security measures to safeguard sensitive health data while tackling scalability issues in real-time settings. The proposed system, hChain 4.0, employs a permissioned blockchain to provide a secure and scalable data infrastructure designed to fulfill these needs. This stands in contrast to conventional systems, which are vulnerable to security flaws or rely on public blockchains, constrained by scalability and expense. The proposed approach introduces a high-privacy method in which health data are encrypted using the Advanced Encryption Standard (AES) for time-efficient encryption, combined with Partial Homomorphic Encryption (PHE) to enable secure computations on encrypted data, thereby enhancing privacy. Moreover, it utilizes private channels that enable isolated communication and ledger between stakeholders, ensuring robust privacy while supporting collaborative operations. The proposed framework enables anonymized health data sharing for medical research by pseudonymizing patient identity. Additionally, hChain 4.0 incorporates Attribute-Based Access Control (ABAC) to provide secure electronic health record (EHR) sharing among authorized parties, where ABAC ensures fine-grained permission management vital for multi-organizational healthcare settings. Experimental assessments indicate that the proposed approach achieves higher scalability, cost-effectiveness, and validated security.
	\end{abstract}


	\section{Introduction}
	
The rapid advancement of smart healthcare technologies has led to a significant increase in the use of devices such as smartwatches and wearable medical appliances. These devices are beneficial for patient health monitoring, data analysis, aiding in the prediction and prevention of health risks \cite{Intro_IoMThighusage}. However, these devices have limited capabilities and low computational power, thus they are more vulnerable to security issues \cite{intro_IoMTsecurity}. Traditional healthcare systems use centralized infrastructure for EHR management and storage, which required enhanced security while handling sensitive information regarding a patient's health. Furthermore, traditional systems are inefficient in handling the lengthy processes of data transfer and EHR management \cite{lengthyTradtional}.
	
\begin{figure*}[htbp]
		\centering
		\includegraphics[width=0.99\textwidth]{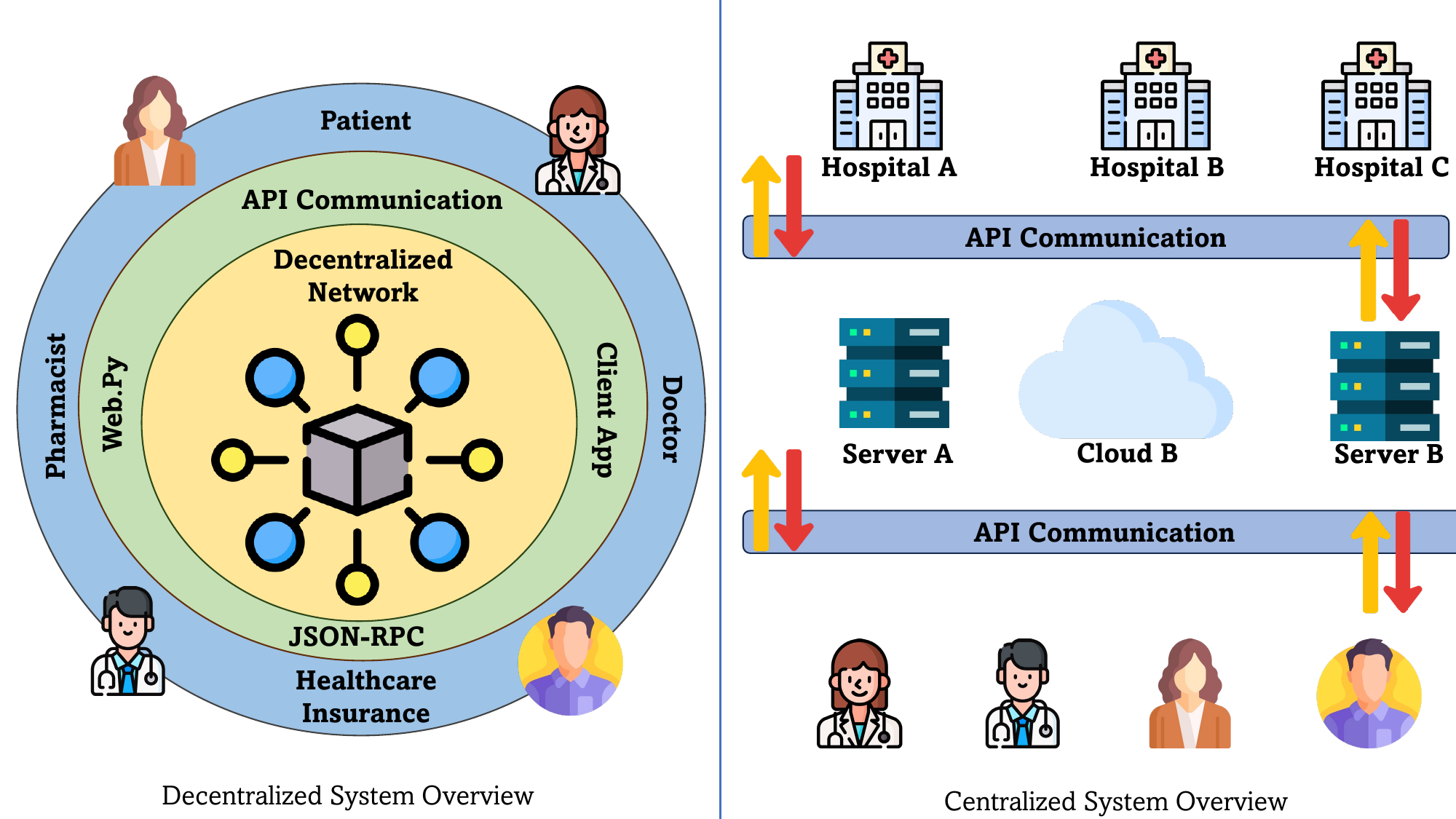}
		\caption{Centralized And Decentralized System Overview.} 
		\label{FIG:CentandDecent} 
	\end{figure*}

	Transitioning from centralized to decentralized data infrastructures addresses the limitations of centralized system. Figure~\ref{FIG:CentandDecent} compares centralized and decentralized systems. In decentralized systems, all stakeholders share a unified infrastructure and data standard, enhancing transparency, patient-centricity, and fault tolerance. Meanwhile, centralized environments rely on isolated servers and varying standards, complicating record-sharing and risking single points of failure. The blockchain, as a distributed ledger technology, ensures data security with its robust security mechanisms. A smart contract is self-executing code that enables secure agreements and automates processes between parties. Blockchain and smart contracts enhance the security and reliability of EHR management while reducing human intervention by automated operations \cite{Intro_BCSC}. Blockchain and smart contracts offer a transformative approach to addressing traditional healthcare challenges.
	
The proposed permissioned blockchain framework utilizes blockchain for EHR and participant management, employing smart contracts to facilitate automated interactions between entities securely. The proposed approach, hChain 4.0, utilizes ABAC to grant secure sharing of EHRs to all authorized participants within the network via a smart contract, fully controlled by the patient. Due the sensitive nature of patient data, a permissioned blockchain is utilized to facilitate restricted access strictly for authorized stakeholders, hence assuring enhanced privacy and security in contrast to a public blockchain, which enables potential access by any individual. It also incorporates PHE for scenarios where research institutes must compute on encrypted EHR data without revealing the underlying information.
	
	In addition to permissioned blockchain security and privacy, hChain 4.0 employs AES-256 encryption to provide another level of privacy for participants. Moreover, public blockchains are associated with higher fees and scalability limitations, whereas hChain 4.0 offers higher scalability without imposing transaction fees. Additionally, channels are utilized to enhance privacy among participating entities by ensuring data confidentiality and isolation from other participants, enabling more efficient data management.
	
	The rest of this paper is organized as follows. The novel contributions of the proposed hChain 4.0 framework are presented in Section \ref{Novel}. Related prior research is discussed in Section \ref{Related}. A detailed architectural overview of hChain 4.0 is provided in Section \ref{Arch}. The proposed algorithms for hChain 4.0 are explained in Section \ref{Algorithm}. Experimental results validating the system are presented in Section \ref{Result}, and finally, the conclusion and future research directions are discussed in Section \ref{Conclusion}.
	\section{Novel Contributions}
	\label{Novel}
	This section outlines the challenges addressed by hChain 4.0 and its innovative contributions in enhancing permissioned blockchain EHR management, focusing on improved privacy, scalability, and reliability.
	\subsection{Problems Addressed}
	
	\begin{itemize}
		\item \textbf{Centralized Limitations \& Data Ownership:} Centralized healthcare systems are prone to single points of failure and lack robust data availability. Inefficient data transfer also delays critical processes, highlighting the need for more effective ownership and control mechanisms.
		\item \textbf{Security Challenges in IoMT Devices:} IoMT devices have limited computational and storage capabilities, making them vulnerable to a variety of security threats.
		\item \textbf{Scalability and Privacy in Public Blockchains:} Public blockchains often have limited throughput, reduced data privacy, high transaction fees, and added latency from consensus protocols, making them less suitable for large-scale, real-time healthcare scenarios.
		\item \textbf{Access Control \& Collaborative Research:} Traditional healthcare systems isolate data across different stakeholders, hindering collaboration and complicating reliable information sharing or validation—especially in multi-organizational research settings.
		
	\end{itemize}
	
	\subsection{Novelty of the Proposed Solution hChain 4.0}
	
	\begin{itemize}
		\item \textbf{Permissioned Blockchain \& Private Channels:} Leverages a permissioned ledger to boost scalability and employs isolated channels for secure communication, ensuring data confidentiality.
		\item \textbf{AES \& PHE:} Integrates AES for time-efficient encryption and adopts PHE to enable secure computations on encrypted data, further enhancing privacy in sensitive healthcare scenarios.
		\item \textbf{ABAC:} Uses chaincode-based ABAC to enforce fine-grained permissions, granting patients robust control over EHR sharing.
		\item \textbf{Smart Contract Automation:} Streamlines stakeholder interactions, minimizes manual intervention, and ensures secure data handling through self-executing contracts.
		\item \textbf{Cost Efficiency \& Anonymized Sharing:} Mitigates the high fees typical of public blockchains and supports anonymized data exchange for research, protecting patient identities.
		\item \textbf{Patient-Centric Interoperability:} Avoids lengthy data transfers by allowing patients direct control, ensuring all authorized parties can access necessary records across organizational boundaries.
	\end{itemize}

	\section{Background}
	
	\subsection{Internet of Medical Things (IoMT)–Based Medical Cyber-Physical Systems (MCPS)}
	
The Internet of Medical Things is often seen as the core of smart healthcare. It connects devices, actuators, and patient physiology to create a Medical Cyber Physical System (MCPS) \cite{MCP_IoMT}. In this system, wearable or implantable devices collect real time data, such as numbers or images, and AI analyzes these data. Based on the analysis, the AI can adjust medication flow or device settings to improve patient care. When hospitals use MCPS, they become smart hospitals, which improves healthcare services, reduces manual tasks, and lowers the workload on practitioners. 

\begin{figure*}[htbp]
	\centering
	\includegraphics[width=0.99\textwidth]{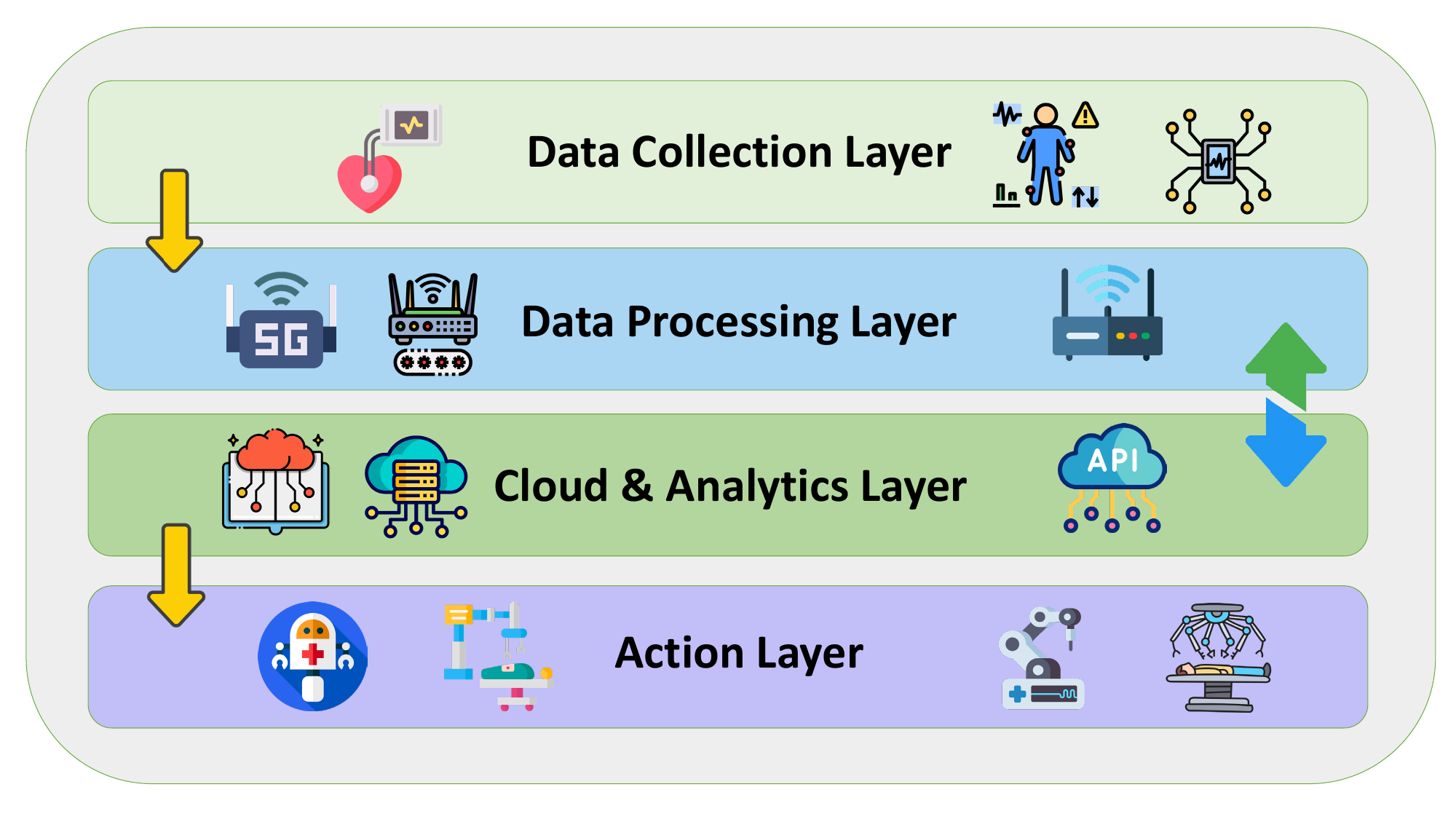}
	\caption{Layered View of IoMT-CPS.} 
	\label{FIG:IoMTCPS} 
\end{figure*}

For example, sensors can be worn by the patient for real time monitoring, and an insulin pump delivers medication to the patient. These devices connect to the Internet and analyze incoming data, whether it is numerical or image based as depicted in Figure \ref{FIG:IoMTCPS}. Based on the analysis, the insulin pump can adjust the medication flow by increasing or decreasing it, or it can alert practitioners if human intervention is required. All of these actions occur through an automated system. An IoMT based MCPS can be divided into multiple layers that handle each stage of this process.

\begin{itemize}
	\item \textbf{Data Collection Layer} This layer consists of various sensors and devices that surround the patient to collect health data \cite{iot_health}. Examples include monitors for heart rate and glucose levels in real time. Each device collects its measurements and sends them to an edge device for further processing and aggregation, which helps form a comprehensive health profile for the patient.
	\item \textbf{Data Processing Layer} In this layer, data originating from multiple sources in real time is received and processed by devices with higher computational capabilities—often referred to as edge devices. Because the initial sensors may have limited resources, tasks such as data formatting, encryption, and authentication are offloaded to these more capable edge nodes \cite{IoT2}. Once the data are properly secured using methods like digital certificates or signatures, the edge devices transmit the information to the cloud infrastructure. This setup not only ensures robust data security but also optimizes the overall system performance by distributing computational responsibilities efficiently.
	\item \textbf{Cloud Layer} This layer primarily provides services through application programming interfaces (APIs) and is responsible for the secure storage and management of data. Owing to its well-established, scalable infrastructure, the cloud layer can support multiple real-time services, making it especially suitable for smart healthcare applications \cite{cloud_IoT}. Upon receiving data from edge devices, the cloud employs authentication methods (e.g., digital signatures) to verify the integrity of the incoming information. Once validated, the data undergoes the required processes—such as advanced analytics, long-term storage, or secure sharing with authorized stakeholders. Finally, if necessary, the cloud relays feedback or alerts to users and relevant healthcare personnel, facilitating real-time patient monitoring and informed clinical decision-making.
	\item \textbf{Analytics Layer} This layer may be deployed at the edge in the processing layer or within the cloud, depending on specific application requirements such as latency, resource constraints, and scalability \cite{edge_cloud_analysis}. It continuously receives data in real time and carries out various analyses according to the system’s needs. These analyses may include predictive modeling, for instance forecasting resource utilization or applying artificial intelligence techniques, to enhance reliability and accuracy. By employing rigorously trained models, this layer facilitates data-driven decision-making in smart healthcare systems, leading to improved patient care and outcomes.
	\item \textbf{Action Layer} This layer carries out the decisions generated by AI models or threshold-based mechanisms. Actions may include regulating the flow of an insulin pump, adjusting a hospital bed for patient comfort or safety, or alerting nearby practitioners in critical situations. In some cases, robotic systems with mechanical actuators can be employed to respond to physical disturbances or perform other automated tasks. By integrating these capabilities, the Action Layer enables real-time responsiveness and closed-loop control in IoMT-based cyber-physical systems, ultimately enhancing patient care and improving overall system reliability.
\end{itemize}

\subsection{Blockchain Applications in Smart Cities}
\begin{figure*}[htbp]
	\centering
	\includegraphics[width=0.99\textwidth]{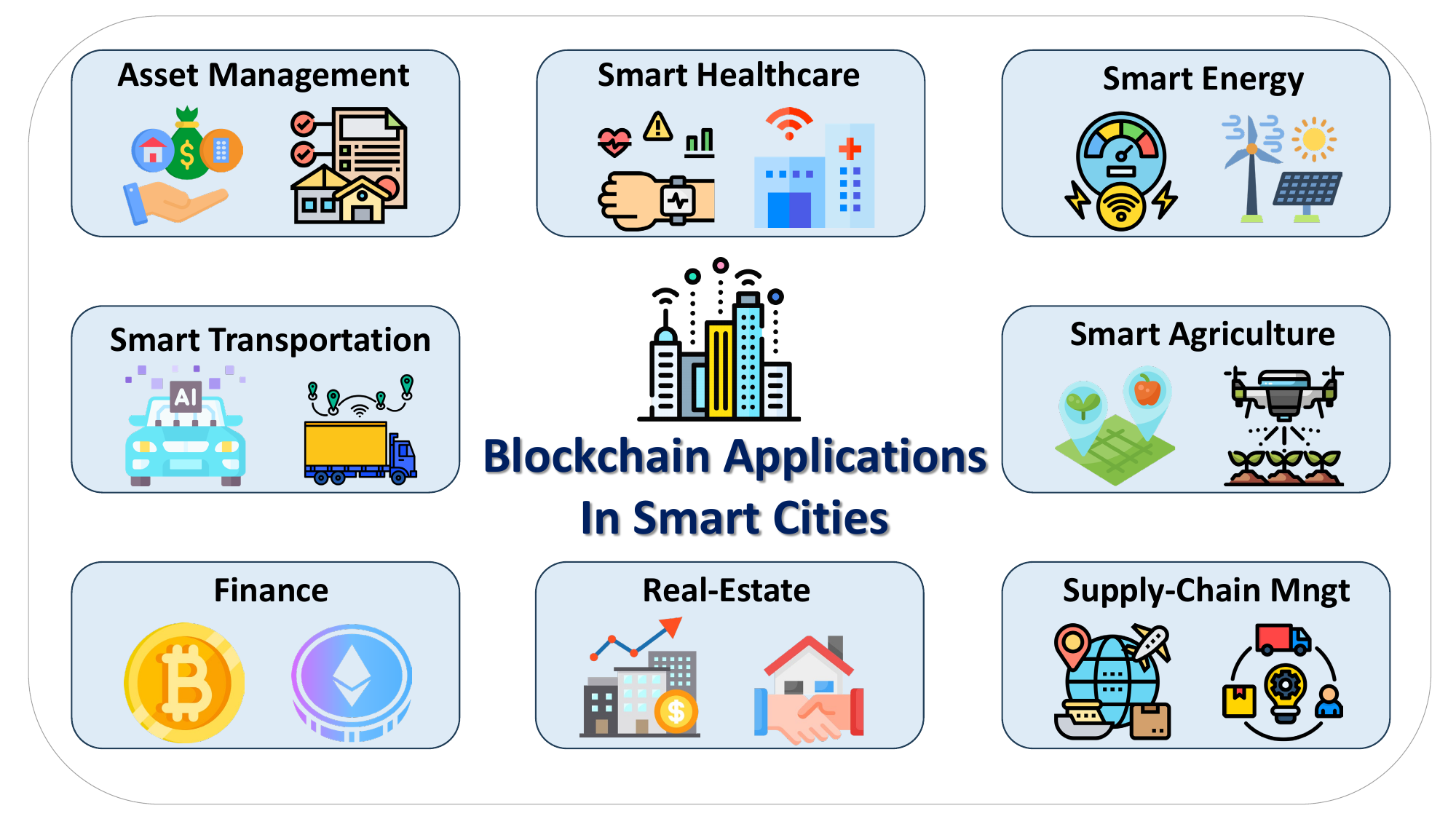}
	\caption{Blockchain Technology Applications In Smart Cities.} 
	\label{FIG:BCApp} 
\end{figure*}

As illustrated in Figure \ref{FIG:BCApp}, blockchain technology is a promising approach for improving smart city systems. Blockchain can be applied in many smart city fields such as healthcare, transportation, supply chain management, and energy, helping to improve each domain. The following subsections explain how blockchain can benefit these domains by outlining its key advantages and practical uses.

\subsubsection{Blockchain in Smart Healthcare}

Blockchain technology offers significant improvements to various aspects of current healthcare systems \cite{BC_Intro1,BC_intro3}. It provides enhanced security, automation, and data integrity—factors that are critical for sensitive medical records and for ensuring patient safety \cite{BC_pros}. Patient healthcare data, for example, are highly confidential and require both protection and controlled sharing across different stakeholders. Blockchain-based Electronic Health Record  management addresses these requirements by maintaining immutable patient data, while also enabling efficient multi-party collaborations. Several studies explore blockchain as a standalone solution or in combination with other technologies to ensure transparency,  integrity or enhancing system security of medical data \cite{BC_App_Security, BC_App_auth_sc,BC_App_FL,BC_App_SC_ACM,BC_App_SC_CE,BC_App_genedata}.

In addition to EHR management, blockchain improves supply chain operations in healthcare. By recording each step of the distribution process, it protects end-users from counterfeit drugs, which can have severe repercussions on patient health. Smart contracts allow patients and medical personnel to verify medications easily, thereby strengthening public trust in the pharmaceutical supply chain. Moreover, the automation of health insurance procedures benefits from blockchain’s tamper-resistant and auditable record-keeping, reducing claims processing time and making contractual agreements between insurers, providers, and patients clearer and more secure.

Blockchain also helps with clinical trials, guaranteeing accurate data management and real-time patient monitoring. In multi-institution healthcare networks, it eliminates duplicate records and reduces the complexity of transferring patient information between different centers—especially relevant for urgent cases. By preserving data integrity, researchers and medical practitioners can rely on secure data to make informed decisions regarding patient care. 

Below are key advantages of using blockchain in smart healthcare:
\begin{itemize}
	\item \textbf{EHR Management}
	\item \textbf{Patient-Centric Control over Records}
	\item \textbf{Transparency in Health Insurance}
	\item \textbf{Improved Clinical Trials}
	\item \textbf{Secure Real-Time Patient Monitoring}
	\item \textbf{Data Standardization}
	\item \textbf{Data Integrity for Researchers}
	\item \textbf{Reducing Counterfeits}
\end{itemize}

\subsubsection{Finance}
In the financial sector, blockchain has gained substantial traction through platforms such as Bitcoin and Ethereum. Major studies investigate the role and impact of blockchain in the finance domain \cite{BC_App_Fin1,BC_APp_Fin2,BC_APp_Fin3}. One of its primary advantages lies in reducing transaction costs, as it minimizes reliance on third-party intermediaries like banks \cite{BC_Fin_cost}. Additionally, blockchain expedites global money transfers by maintaining a distributed ledger that updates in near real time. This decentralized structure also strengthens security, since no single entity can alter or manipulate transaction records without consensus. Blockchain smart contracts further automate financial processes by establishing predefined rules that execute transactions once certain conditions are met. Moreover, blockchain’s immutable and transparent nature enhances auditing practices, while its global reach streamlines cross-border transactions. Finally, blockchain technology supports micropayment solutions, making smaller transactions economically viable and accessible worldwide.

Below are key advantages of adopting blockchain in finance:

\begin{itemize}
	\item \textbf{Lower Transaction Costs}
	\item \textbf{Faster Global Money Transfers}
	\item \textbf{Enhanced System Security}
	\item \textbf{Automated Processes via Smart Contracts}
	\item \textbf{Improved Auditability}
	\item \textbf{Simplified Cross-Border Transactions}
\end{itemize}

\subsubsection{Asset Management}

Tokenization stands as a key innovation in blockchain technology, allowing physical assets to be represented digitally \cite{Token_def}. This process simplifies trading by reducing reliance on intermediaries, enhances transparency, and provides improved audibility of transactions. Large or traditionally illiquid assets, such as real estate or artwork, can be divided into smaller tradable units, thus increasing market liquidity. Furthermore, automated processes via smart contracts streamline compliance with regulations, including Know Your Customer (KYC) protocols, and help safeguard against fraud. By lowering administrative costs and offering immutable proof of ownership, blockchain fosters greater confidence among investors and paves the way for flexible, fractional ownership of digital assets. Various studies have explored the application of blockchain technology for asset management across diverse fields \cite{BC_App_Asset1,BC_App_Asset2,BC_App_Asset3}.
Below are key advantages of adopting blockchain in asset management:
\begin{itemize}
	\item \textbf{Increased Liquidity Through Digital Asset Representation}
	\item \textbf{Secure Automation Using Smart Contracts}
	\item \textbf{Reduced Reliance on Third Parties and Associated Costs}
	\item \textbf{Efficient Asset Tracking and Transfer}
	\item \textbf{Enhanced Investor Confidence}
	\item \textbf{Fractional Ownership of Digital Assets}
\end{itemize}

\subsubsection{Real Estate}

Blockchain technology has the potential to transform various aspects of real estate by increasing transparency, trust, and efficiency \cite{BC_App_re}. Through tokenization, blockchain enables a digital representation of property ownership, allowing stakeholders to verify and transfer assets in a secure and transparent manner. Because of its immutable ledger, blockchain significantly reduces the risk of fraud, as any attempt to alter records would be immediately evident. In addition, smart contracts can automate the transfer of ownership, which simplifies administrative processes and decreases dependency on third-party intermediaries. Lower intermediary involvement reduces overall costs, including administrative fees, and broadens market access. Non-Fungible Tokens (NFTs) can further support fractional ownership of high-value assets \cite{BC_App_rs_NFT_Fraction}, offering flexible investment options for a wider audience. Extensive academic studies have examined blockchain technology in real estate to address current system deficiencies or enhancing existing solutions \cite{BC_App_rs11,BC_App_rs22,BC_App_rs33}.

Below are key advantages of adopting blockchain in real estate:

\begin{itemize}
	\item \textbf{Enhanced Trust Among Stakeholders}
	\item \textbf{Simplified Ownership Verification}
	\item \textbf{Fraud Reduction}
	\item \textbf{Automated and Accelerated Property Transfers}
	\item \textbf{Lower Intermediary Costs}
	\item \textbf{Fractional Ownership through Tokenization}
	\item \textbf{Auditability and Transparency of Property Histories}
\end{itemize}

\subsubsection{Smart Energy}

Blockchain technology and smart contracts can advance smart energy initiatives. One notable application is peer-to-peer energy trading. Users with surplus solar or other renewable resources can exchange power directly without centralized intermediaries. This decentralized approach enhances billing transparency and allows automated pricing based on demand and supply. Smart energy can have secure, transparent traceability and efficient pricing through blockchain’s decentralized, tamper-proof mechanism \cite{BC_App_SE_cit}. Transparent traceability of energy sources increases trust among stakeholders and fosters a more efficient energy market. Several works \cite{BC_App_SE1,BC_App_SE2,BC_App_SE3,BC_App_SE4,BC_App_SE5} highlight blockchain’s advantages for secure, transparent, and privacy-focused smart energy systems.

Below are key advantages of adopting blockchain in smart energy:

\begin{itemize}
	\item \textbf{Enabling P2P Energy Trading}
	\item \textbf{Providing Transparent Billing}
	\item \textbf{Encouraging Renewable Energy Adoption}
	\item \textbf{Automated Price Adjustments}
	\item \textbf{Lowering Administrative and Billing Costs}
\end{itemize}

\subsubsection{Smart Agriculture}

Researchers have investigated the use of blockchain in agriculture to leverage its blockchain mechanisim for data quality or integration with IoT to enhance overall framework \cite{BC_App_SA1,BC_App_SA2,BC_App_SA3}. One prominent application lies in tracing products from the farm to the marketplace, ensuring that each item’s origin is accurately documented. This transparency bolsters consumer trust, since the authenticity and quality of produce can be verified on a tamper-resistant ledger. Blockchain-based systems also minimize fraud by recording product details and enabling consumers to validate these records via smart contracts. Furthermore, they facilitate real-time monitoring of farms, improving oversight of crop conditions and guaranteeing that produce meets health and safety requirements. Finally, blockchain supports secure, cross-border data sharing among stakeholders, fostering cooperation and confidence throughout the supply chain.

Below are key advantages of adopting blockchain in smart agriculture:
\begin{itemize}
	\item \textbf{Tracing Produce from Farm to Market}
	\item \textbf{Reducing Food Fraud}
	\item \textbf{Reliable Farm Monitoring}
	\item \textbf{Secure Cross-Border Data Sharing}
	\item \textbf{Real-Time Produce Traceability}
\end{itemize}

	\subsubsection{Supply Chain Management}

	Blockchain technolgy provides serveral advtanges and services to supply chain applications \cite{BC_App_SCM1,BC_App_SCM2}. It ensures the goods and products and medicince are delivered safely through utilizting IoT devices along with blockchain for ensuring the tracking with security and providing integrity of the data generated from CPS devices that monitorrs the goods in real-time. It also providing a way to easiliy identify the items that are on recall with trustworthy system. Another applications can be goods or food soruce verification through using smart contract and identifical ID for each items and storing the details of the product on blockchain so easily can be retirived through smart contract in secure manner. It also providing transparency due to the data is distriubted in decentralied manner for reulatory compiance and faclitate the auditiability due to the immutable ledger. It also can be used to provbied payment based on the delivey status in secure manner with condidentally the system security.  It also avoids addtional works associated with data entry each time accross the border.
	
	Below are key advantages of adopting blockchain in supply chain management:
	
	\begin{itemize}
		\item \textbf{Tracing Produce from Farm to Market}
		\item \textbf{Reducing Food Fraud}
		\item \textbf{Reliable Farm Monitoring}
		\item \textbf{Secure Cross-Border Data Sharing}
		\item \textbf{Real-Time Produce Traceability}
	\end{itemize}
	
	\subsubsection{Smart Transportation}

There are limitations in transportation existing systems that can be addressed with blockchain technology and smart contracts \cite{BC_App_T_limitation}. A decentralized architecture provides more transparency and security while improving data integrity and reducing reliance on central authorities. This approach is beneficial for use cases such as vehicle-to-vehicle communication \cite{BC_App_TV2V} where trustworthy data exchange is essential. By automating rules and processes with smart contracts the system reduces errors or malicious interference and lowers operational costs. One of the key benefits of integrating blockchain into smart transportation is the ability to digitize ownership through tokenization \cite{BC_App_TokenTN}, making it easier and more secure to transfer vehicle titles. This reduces both the time and operational expenses normally associated with changing ownership. Through this process, stakeholders in the transportation sector gain a more streamlined experience, as tokenized assets allow precise tracking, reduced paperwork, and greater transparency overall. In addition, blockchain solutions can enable automated toll payments and enhance accountability in transportation networks. They also streamline vehicle recall procedures by making it easier to identify and track cars through a tamper-proof record of ownership and service history.

Below are key advantages of adopting blockchain in smart transportation:

\begin{itemize}
	\item \textbf{Simplifies Recall Procedures} 
	\item \textbf{Secures V2V Communications} 
	\item \textbf{Facilitates Ownership Transfer} 
	\item \textbf{Simplifies Tollway Payments} 
	\item \textbf{Ensures Vehicle History Integrity} 
	\item \textbf{Improves Accountability} 
\end{itemize}

	\section{Smart Healthcare: Challenges}
	
\begin{figure*}[htbp]
	\centering
	\includegraphics[width=0.99\textwidth]{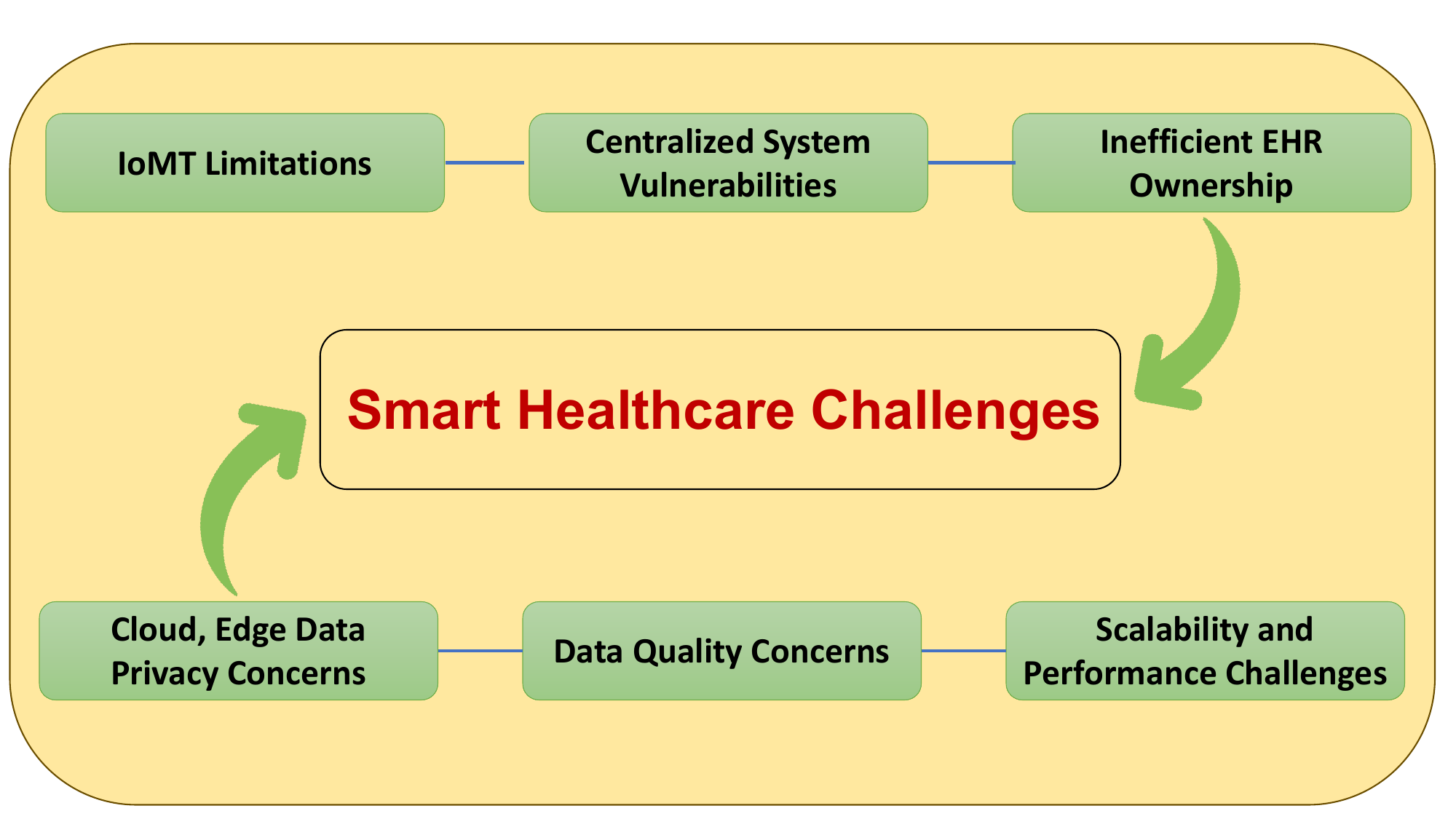}
	\caption{Smart Healthcare Challenges.} 
	\label{FIG:SHChallenges} 
\end{figure*}

Smart healthcare uses multiple technologies to deliver advanced medical services. However each component has potential drawbacks as depicted in Figure \ref{FIG:SHChallenges}. IoMT devices often have limited resources, creating security and performance challenges \cite{ioMT-Seucrity1}. Storing data in the cloud can raise privacy concerns and lead to breaches if not managed securely \cite{Cloud_Security_concern}. Although edge devices can mitigate IoMT constraints by performing local processing, they may also introduce additional vulnerabilities and expand the system’s attack surface.  In addition, privacy and security of EHRs are critical. Any manipulation or loss of integrity can lead to severe consequences. Interoperability among heterogeneous systems remains a major challenge, since the absence of unified standards hinders smooth data exchange and complicates analysis or retrieval. Scalability and performance may also degrade when the system grows while preserving confidentiality, integrity, and availability. Furthermore, data quality is a key concern in healthcare  \cite{data_quality}, especially when using real-time patient monitoring or generating health data for research purposes. The following are some of the challenges of smart healthcare:
\begin{itemize}
	\item \textbf{Limited IoMT Resources}
    \item \textbf{Centralized System Vulnerabilities}
	\item \textbf{Inefficient EHR Ownership}
	\item \textbf{High Latency in EHR Transfers}
	\item \textbf{Limitations in Access Control Management}
	\item \textbf{Single Point of Failure (SPoF)}
	\item \textbf{Cloud Data Privacy Concerns}
	\item \textbf{Edge Device Vulnerabilities}
	\item \textbf{Privacy and Security of EHRs}
	\item \textbf{Interoperability Among Heterogeneous Systems}
	\item \textbf{Scalability and Performance Challenges}
	\item \textbf{Data Quality Concerns}
	\item \textbf{Interoperability Among Heterogeneous Systems}
	\item \textbf{Scalability and Performance Challenges}
\end{itemize}

	\section{Related Prior Research}
	\label{Related}
	
	Various approaches have been implemented in the healthcare sector to enhance traditional centralized EHR management, focusing on optimizing efficiency, strengthening security, and implementing robust access control mechanisms \cite{RW_Cloud4,RW_Cloud1,RW_Cloud2,RW_cloud3,RW_RDBMS}. However, IoMT devices face different security threats and concerns to be considered in the healthcare domain \cite{CLR}. The approach employs blockchain technology within a privacy-preserving IoMT framework, incorporating both lightweight on-chain and off-chain storage for secure data distribution \cite{LR1}. The authors presented a blockchain-based biometric security architecture for IoMT devices to deliver a decentralized identification solution and robust cryptographic approaches to tackle privacy and authenticity concerns in IoMT \cite{LR2}. This work employs ECDSA-based techniques in conjunction with federated learning and blockchain to address privacy concerns \cite{LR3}. This method combines fog computing and CP-ABE to ensure secure EMR sharing with fine-grained IoMT access control \cite{LR4}. In addition, many studies presented decentralized and scalable EHR management cloud-based approaches to address the security and access control management issues in centralized system. Table \ref{tab:summary_comparison_inverted} illustrates the features and technical comparison of state-of-the-art works and the proposed approach. 

\begin{table*}[htbp]
	\centering
	\caption{Comparative Table for State-of-the-Art Works}
	\label{tab:summary_comparison_inverted}
	\begin{tabularx}{\textwidth}{|l|X|X|X|X|X|X|X|}
		\hline
		\textbf{Work/Paper} & \textbf{Blockchain Type} & \textbf{Encryption Scheme} & \textbf{Isolation / Private Channels} & \textbf{Access Control Mgmt.} & \textbf{ID Removed for Researchers} & \textbf{Scalability} & \textbf{Tx/ Operational Cost} \\
		\hline
Zala, et al. 2022		\cite{RW3} & Cloud-Based & AES-128, Steg. & \ding{55} & Cloud Access & \ding{55} & High & Moderate \\
		\hline
Katoch, et al. 2023		\cite{RW6} & Permi\-ssioned & N/A & \ding{55} & N/A & \ding{55} & High & Low \\
	\hline
Haddad, et al. 	2023	\cite{RW2} & Permi\-ssioned Infra. & Session Keys (IPFS) & \ding{55} & RBAC & \ding{55} & Moderate & Moderate \\
		\hline
Pandit, et al. 2024		\cite{RW5} & Public & AES-128 & \ding{55} & ABAC & \ding{55} & Moderate & Moderate \\
	\hline
Ghugare, et al. 2024	\cite{RW1} & Private & AES-256 & \ding{55} & RBAC & \ding{55} & Moderate & Moderate \\
		\hline
		\textbf{hChain 4.0} & Permi\-ssioned & AES-256, PHE & \ding{51} & ABAC & \ding{51} & High & Low \\
		\hline
	\end{tabularx}
\end{table*}

	In ref. \cite{RW1}, the authors propose a comprehensive approach for EHR and patient registration that offers secure health data storage while preserving data integrity through the use of data hashes and an automated scheduling mechanism. The system addresses data confidentiality requirements within a private blockchain-based healthcare framework. When patients register, they receive unique private keys alongside access control parameters, all managed and stored within a smart contract. To protect sensitive information, the authors employ advanced encryption methods, storing health data off-chain in a distributed IPFS environment, with only the data hashes maintained on-chain to ensure data integrity and improve scalability. The smart contract utilizes a basic role-based access control model and supports automated processes such as appointment scheduling, reducing administrative overhead. As a result, sensitive health data is encrypted and transferred to cloud storage, reinforcing privacy, security, and scalability.

	The authors in \cite{RW2} introduce a patient-centered, blockchain-based EHR management system known as PCEHRM that aims to enhance the current healthcare infrastructure by enabling more secure data sharing and storage of critical health information without relying on centralized methods. This solution integrates Ethereum blockchain with IPFS to improve scalability and reduce costs, incorporating smart contracts as well. The framework allows stakeholders such as patients, researchers, and laboratories to access and manage EHRs securely and efficiently. In particular, the patient-centric access control mechanism governs permission grants and revocations to protect data privacy. The implementation employs Truffle to deploy and test smart contracts and utilizes Web3 in a Windows environment to confirm the feasibility and effectiveness of the system. Evaluations of storage costs at varying execution times with different peers and document sizes demonstrate that the proposed approach is both scalable and practical.

	The authors in \cite{RW3} proposed approach is called PRMS system utilizes a cloud-based communication approach that involves three main stakeholders: patients, doctors, and relatives, functioning within an e-health environment hosted on services such as AWS and GCP. A window-based data collection strategy records both textual and patient information, thereby improving real-time monitoring and facilitating secure allocation of data across geographically distributed centers. The architecture is grounded in five core processes: authentication, cryptographic encryption, steganography-based data hiding, access control, and a hybrid AES-128 and LSB framework to provide two-layer security in a third-party cloud. This integration enables users to securely upload, access, and share patient records while allowing fine-grained privileges. The design supports activities such as consulting with physicians, scheduling appointments, and uploading medical reports, thereby safeguarding data integrity and patient privacy. In addition, it compares throughput metrics against existing blockchain-based solutions.

	The authors in \cite{RW5}  introduce a patient-centric Electronic Health Record management system built on Ethereum. The proposed architecture involves five entities: EHR users, patients, IPFS, an EHR File Manager, and the Ethereum blockchain, all of which collectively ensure secure and transparent data handling. By leveraging decentralized ledger capabilities and smart contracts, the system applies fine-grained access control through Cipher-text Policy Attribute-Based Encryption, implements encryption mechanisms for data confidentiality, and supports efficient processes for uploading and viewing EHRs. Key features include validating and verifying access permissions, storing and retrieving encrypted records through IPFS, and embedding security attributes within the blockchain. Through various operations such as user registration, granting EHR access, uploading records, and viewing files, this approach demonstrates how blockchain can optimize EHR sharing and management while mitigating issues like unauthorized access, single points of failure, and limited auditability.

	In ref. \cite{RW6}, the authors adopt a permissioned blockchain architecture grounded in Hyperledger Fabric to streamline e-healthcare transactions, thereby strengthening security, confidentiality, and overall efficiency. They also address significant use cases such as clinical trials, drug supply chain management, EHR administration, insurance claims, and medical device tracking. To evaluate performance, the authors investigate different network configurations comprising one organization with one peer, two organizations with one peer, and two organizations with two peers, and they employ Hyperledger Fabric with the PBFT consensus mechanism to boost efficiency and reduce latency while private channels are not discussed. The results underscore the capacity of blockchain-based solutions to considerably enhance trust and reliability in healthcare data exchanges while safeguarding patient privacy, which is essential given the sensitive nature of healthcare information.

	
	In our previous works, collectively referred to as hChain 1.0, 2.0, and 3.0, we explored various blockchain-based EHR management approaches \cite{hchain,hchain2,hchain3}. In hChain 1.0 \cite{hchain}, we employ a public blockchain with location-based authentication to enhance security, whereas hChain 2.0 \cite{hchain2} integrates IPFS off-chain storage to improve scalability and cost-efficiency, and utilizes Long Short-Term Memory (LSTM) for fall detection. In addition, hChain 3.0 \cite{hchain3} leverages cloud-based storage for EHR while retaining only the hash on the blockchain, using RBAC with multi-layer authentication. Table~\ref{tab:hchain_variants_inverted} details the progression from these earlier versions toward hChain 4.0’s permissioned architecture, which further enhances scalability, privacy, and real-time responsiveness.
	
	\begin{table*}[htbp]
		\centering
		\caption{Comparative Table for hChain Research}
		\label{tab:hchain_variants_inverted}
		\begin{tabularx}{\textwidth}{|l|X|X|X|X|X|X|X|}
			\hline
			\textbf{Work/Paper} & \textbf{Block\-chain Type} & \textbf{Conse\-nsus Mechanism} & \textbf{Access Control Mgmt} & \textbf{Encry\-ption} & \textbf{Isolation} & \textbf{Scalability} & \textbf{Cost} \\
			\hline
			Alruwaill, et al. 2023	 (hChain)	\cite{hchain} & Public & PoS & RBAC & Symm\-etric & \ding{55} & Low & High \\
			\hline
			Alruwaill, et al. 2023	(hChain 2.0) \cite{hchain2} & Public & PoS & RBAC & Asymm\-etric & \ding{55} & Moderate & Moderate \\
			\hline
			Alruwaill, et al. 2024	(hChain 3.0) \cite{hchain3} & Public & PoS & RBAC & Symm\-etric & \ding{55} & Moderate & Moderate \\
			\hline
			\textbf{hChain 4.0} & Permi\-ssioned & Raft & ABAC & Symm\-etric & Proposed & Highest & Low \\
			\hline
		\end{tabularx}
	\end{table*}
	
	However, hChain 4.0 implements a permissioned blockchain architecture with the Raft consensus mechanism, thereby enhancing throughput, scalability, and cost efficiency. It employs AES-256 encryption to minimize encryption time while multiple chaincodes that omit patient identifiers to facilitate medical research under robust data privacy constraints and encrypted EHR management. Furthermore, secure channels and separate ledgers across participating organizations reinforce confidentiality by isolating sensitive EHR operations among diverse stakeholders. Additionally, ABAC provides fine-grained permission management for EHR data. This ensures that only authorized parties can view or modify records. Meanwhile, PHE enables limited computations on encrypted patient information. This preserves patient confidentiality by preventing any disclosure of sensitive information.

	Table \ref{tab:consensus_comparison} provides a brief comparison of Raft and PBFT, indicating that Raft adopts a more straightforward methodology appropriate for crash-only failures, while PBFT supports broader fault models by tolerating malicious behavior. Thus, Raft is generally more efficient and simpler for smaller, trusted environments, whereas PBFT is more suitable for situations where nodes may behave maliciously.

\begin{table*}[htbp]
	\centering
	\caption{Comparison of Permissioned Consensus Protocols: Raft and PBFT}
	\label{tab:consensus_comparison}
	\begin{tabularx}{\textwidth}{|X|X|X|}
		\hline
		\textbf{Parameter}
		& \textbf{Raft}
		& \textbf{PBFT}
		\\ \hline
		
		\textbf{Mechanism}
		& Leader-based replication
		& Byzantine multi-phase
		\\ \hline
		
		\textbf{Complexity}
		& O(n)
		& O(n\textsuperscript{2})
		\\ \hline
		
		\textbf{Max Number of Fault-Tolerant Nodes}
		& \(2f + 1 \le n\)
		& \(3f + 1 \le n\)
		\\ \hline
		
		\textbf{Fault Tolerance}
		& Crash fault tolerant
		& Byzantine-fault tolerant
		\\ \hline
		
	\end{tabularx}
\end{table*}

\section{Architectural Overview of hChain 4.0}
\label{Arch}

\begin{figure*}[htbp]
		\centering
		\includegraphics[width=0.99\textwidth]{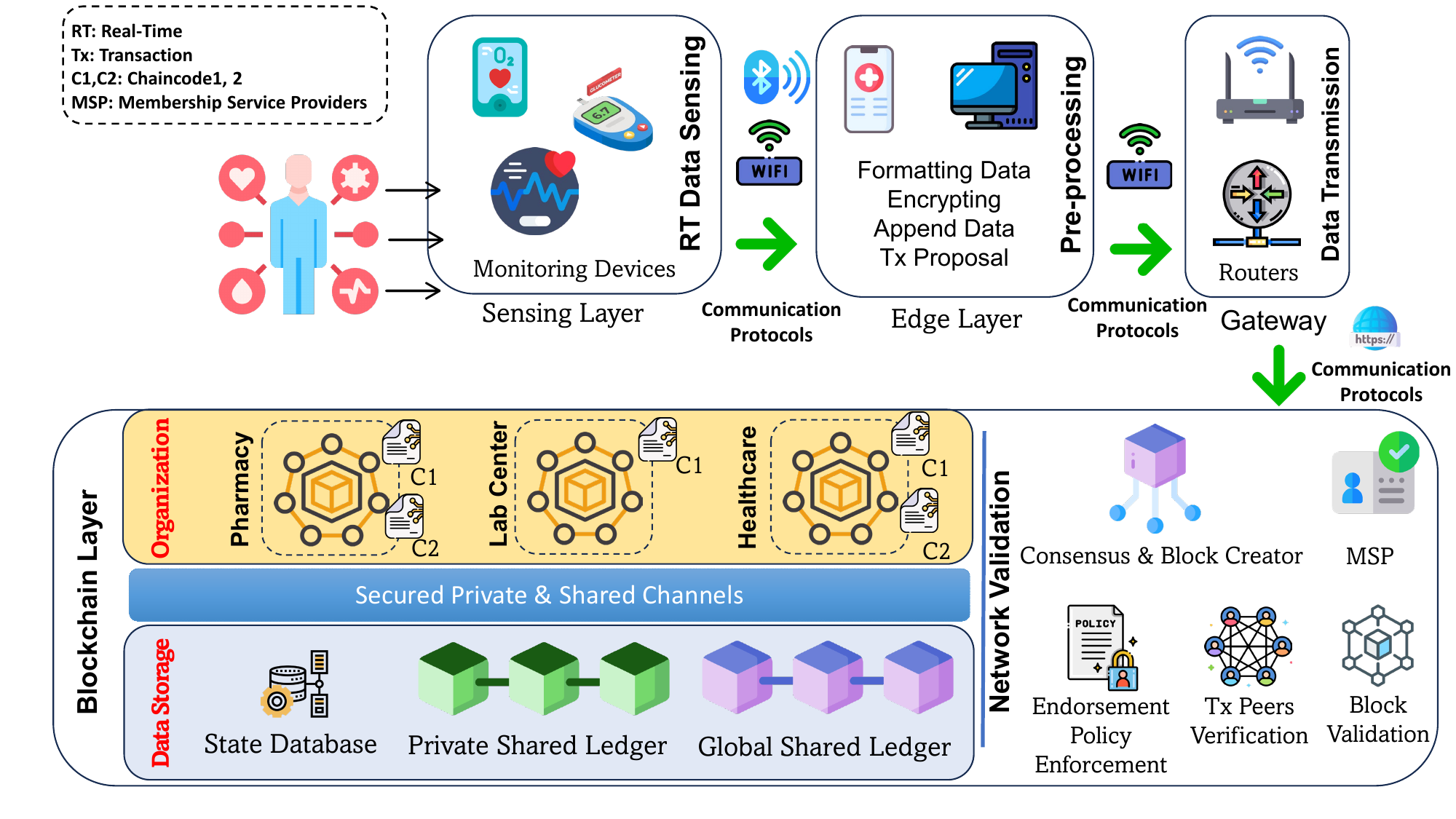}
		\caption{hChain 4.0 Framework Architecture.} 
		\label{FIG:SysArch} 
	\end{figure*}

The hChain 4 architecture comprises three primary layers as depicted in Figure \ref{FIG:SysArch}, each performing a distinct function to ensure secure, reliable, and efficient data management. The first layer, referred to as the sensing layer, continuously gathers real-time data from wearable devices and forwards these data to the edge layer.

Within the sensing layer, smart devices such as wearables generate physiological data in real time. Each device connects to an edge device, which holds the necessary credentials and configuration details to interact with the blockchain network. Since these sensing devices typically possess limited computational and security capabilities, the data they produce are initially vulnerable. Consequently, the edge layer ensures the confidentiality of these data by applying AES-256 encryption, thus enhancing data privacy in local environments. It also standardizes the data format to facilitate advanced analytics. Once these processes are complete, the edge layer submits the resultant encrypted data to the blockchain network through a dedicated client application.
	
	The client application serves as a communication bridge between the edge layer and the blockchain, submitting transactional proposals and awaiting endorsements from designated peers. Upon receiving sufficient endorsements, the blockchain commits the transactions to its distributed ledger. At the blockchain layer, multiple organizations maintain a shared public ledger while retaining operational independence. Moreover, it establishes separate ledgers and multiple chaincodes are deployed for specialized collaboration. A dedicated research organization is also established to facilitate the secure exchange of EHRs without patient identifiers, removing the need for additional patient actions. Additionally, hChain4 integrates PHE to enable secure computations on encrypted data, letting research entities derive insights without ever revealing raw patient information. This approach further bolsters privacy by ensuring confidential operations remain hidden, even across multi-organizational collaborations. In addition to real-time data management, the chaincodes are able to maintain the patient history and critical operation management with multiple signature approval from different stakeholder to ensure different stakeholder such as healthcare providers to approve the operation before committed and approved. ABAC ensures fine-grained EHR management in hChain4, empowering patients while enabling diverse stakeholders to securely access anonymized or PHE-encrypted records. It is particularly suitable for large, multi-organizational healthcare ecosystems, providing flexible permission structures that accommodate complex roles and policies.
	
	The Raft consensus algorithm is employed to optimize network scalability. Moreover, a transaction is deemed approved only after receiving endorsements from multiple peers across different organizations, thereby ensuring robust validation within the network. To optimize performance, LevelDB is utilized, enabling rapid read and write operations, simplified query mechanisms, and a streamlined data structure for efficient storage and usage. The primary functionalities of the chaincode are detailed in Table \ref{tab:chaincode-functions}.

\begin{table}[htbp]
	\centering
	\footnotesize
	\caption{Chaincode Functionalities Overview}
	\label{tab:chaincode-functions}
	\begin{tabular}{p{3.2cm} p{12.3cm}}
		\toprule
		\textbf{Function Name} & \textbf{Description} \\
		\midrule
		
		\textbf{\texttt{AssignAttributes}} 
		& Sets or updates user attributes under ABAC rules. 
		Only an admin (ledger or MSP-level) can assign roles 
		or other metadata to the target user. \\
		
		\textbf{\texttt{GetUserAttributes}} 
		& Retrieves the ABAC attributes of a specified user from the ledger. 
		Returns a default ``guest'' role if the user record is absent. \\
		
		\textbf{\texttt{CreateEHR}} 
		& Allows a user with \texttt{role=Patient} to create a new 
		encrypted EHR entry. Checks policy, then stores the record 
		under a dedicated key. \\
		
		\textbf{\texttt{QueryEHR}} 
		& Permits a user with \texttt{role=Doctor} 
		(or MSP admin fallback) to retrieve an existing EHR. 
		Enforces ABAC checks before reading ledger data. \\
		
		\textbf{\texttt{ShareEHR}} 
		& Enables a user with \texttt{role=Patient} 
		to share EHR records with another party by writing 
		a lightweight “share” record, granting read permission. \\
		
		\textbf{\texttt{ApproveEHR}} 
		& Allows a \texttt{Doctor} to mark an EHR as “approved” 
		(e.g., for medical decisions). Currently logs an event 
		but can be extended for more complex approvals. \\
		
		\textbf{\texttt{AddEncryptedEHRs}} 
		& Demonstrates homomorphic addition on two EHR ciphertexts. 
		 \\
		
		\bottomrule
	\end{tabular}
\end{table}

	\subsection{Proposed Algorithms for hChain 4.0}
	\label{Algorithm}
	
\begin{figure*}[htbp]
		\centering
		\includegraphics[width=0.99\textwidth]{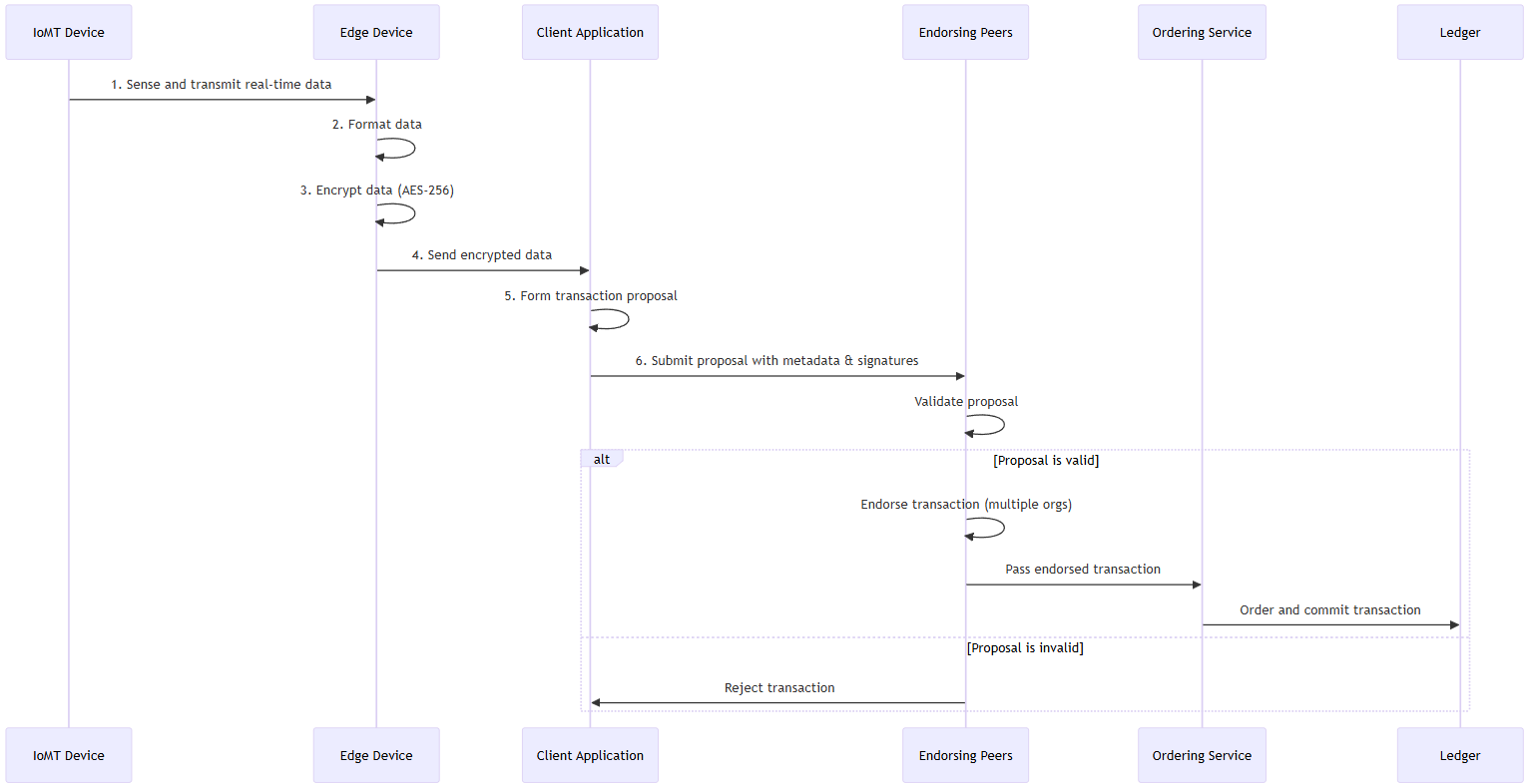}
		\caption{Data Workflow in Sequence Diagram.} 
		\label{FIG:seq} 
	\end{figure*}

	Algorithm \ref{ALG:hChain4_DataFlow} describes the complete hChain 4.0 data flow, from the generation of sensor-derived data to the commitment or rejection of transactions within the blockchain network. In the initial phase, IoMT devices sense data and transmit it in plain text to an edge device. The edge device then preprocesses, reformats, and encrypts the data, forwarding it to the blockchain network through a client application as depicted in \ref{FIG:seq}. Subsequently, the client application constructs a transaction proposal, which is evaluated by endorsing peers from multiple organizations to verify transaction validity. If the transaction is approved, it is committed to the blockchain; otherwise, it is rejected.
	
	\begin{algorithm}[htbp]
		\caption{hChain 4: Data Flow from Sensing to Blockchain Storage}
		\label{ALG:hChain4_DataFlow}
		\begin{algorithmic}[1]
			\REQUIRE
			\begin{itemize}
				\item IoMT Device, Edge device, Valid Account
			\end{itemize}
			\ENSURE
			\begin{itemize}
				\item Transaction approval by multiple organizations
			\end{itemize}
			\STATE \textbf{Sensing Layer}
			\STATE IoMT device senses and transmit real-time data to the edge device.
			\STATE \textbf{Edge Layer}
			\STATE Format data.
			\STATE Encrypt data with AES-256.
			\STATE Send encrypted data using the client application.
			
			\STATE \textbf{Client Application}
			\STATE Form a transaction proposal containing encrypted data, metadata, and signatures.
			
			\STATE \textbf{Blockchain Layer}
			\STATE Endorsing peers validate the proposal.
			\IF{Valid}
			\STATE Peers in different organizations endorse the transaction.
			\STATE Ordering service sequences transactions.
			\STATE Commit transaction to the ledger.
			\ELSE
			\STATE Reject the transaction.
			\ENDIF
		\end{algorithmic}
	\end{algorithm}

	\section{Implementation And Validation}
	\label{Result}
	
	\subsection{Implementation}
	
	Hyperledger fabric 2.5 is utilized as permssioned blockchain and Caliper for testing the network performance. In addition, Python and JavaScript is used to make a test code and interact with the network. The deployment and testing are performed using Mac OS Monterey with 2.2 GHz Quad-Core Intel Core i7 processor and 16 GB 1600 MHz DDR3 memory. Chaincode is written in the Go programming language, as Hyperledger Fabric supports Go for chaincode development.

	\subsection{Validation}
	
	\subsubsection{Time And Cost Analysis}
	The Ethereum public blockchain utilized in hChain, hChain 2.0, and Chain 3.0 presents latency concerns and is less scalable compared to permissioned blockchains. The estimated transaction rate for the Ethereum blockchain is approximately 25 transactions per second (TPS) \cite{EthTPS}.  During the testing instance utilizing a caliper, it demonstrated 483 TPS, which is an increase of 19 times. Furthermore, hChain 4.0 shows improved cost efficiency than \cite{hchain,hchain2,hchain3}. Moreover, AES-256 is employed for its lower computational overhead and faster encryption compared to asymmetric algorithms. As \cite{AES_RSA} shows that AES achieves \(O(m)\) time complexity for an input message of size \(m\) and RSA shows higher complexity \(O(\{ \log n \}^3)\) for the private key and greater processing delays for volume data. This performance gap is crucial for hChain4 which requires minimal latency in encrypting real-time data. Furthermore, hChain4 utilizes LevelDB for its higher throughput compared to CouchDB, a critical factor in a real-time framework. As transaction complexity grows, CouchDB experiences a notable decline in performance, whereas LevelDB remains unaffected \cite{leveldb_couchdb}.
	\subsubsection{Data Privacy and Access Control Management}
	hChain4 employs a permissioned blockchain architecture with channels, allowing organizations to maintain isolated ledgers. Only authorized parties within a channel can view records, preventing sensitive data leaks beyond the intended group. An ABAC mechanism governs permissions at the chaincode and ledger levels, using participants’ roles to determine access. To protect highly confidential data, AES-256 encryption is applied before storing records on the blockchain. Additionally, hChain4 integrates PHE so that research centers can compute on encrypted health data without viewing plaintext, thereby preserving patient privacy while still enabling meaningful analysis. By combining channel-based isolation, ABAC policies, strong encryption, and PHE-driven secure computations, hChain4 maintains robust data confidentiality and privacy for multi-organizational healthcare networks.
	
	\subsubsection{Data Security}
	
	hChain4 addresses each threat with robust security measures. Sybil attacks are prevented by a membership service that demands valid certificates, blocking unauthorized nodes. The immutable, hash-linked ledger structure makes silent data modification impossible and swiftly exposes tampering. Transaction replay attempts are mitigated because endorsing peers reject repeated unique transaction IDs. DDoS threats are mitigated via permissioned membership, channel segmentation, and Raft’s leader-based approach that handles heavy loads. AES encryption and TLS connections also shield data exchanges, reducing the risk of eavesdropping. Together, these measures preserve the confidentiality, integrity, and availability of real-time healthcare data.
	
	\subsection{Scalability}
	Earlier versions of hChain (1.0, 2.0, 3.0) relied on a public PoS Ethereum testnet (Sepolia) with an average confirmation time of 10 seconds per transaction. In contrast, under a 32-worker load, hChain4 achieves around 483.7 TPS throughput. Table~\ref{tab:createEHR_results} shows the detailed benchmark results for the \texttt{CreateEHR} transaction.

\begin{table*}[htbp]
	\centering
	\caption{Benchmark Results}
	\label{tab:createEHR_results}
	\begin{tabularx}{\textwidth}{|X|X|X|X|X|}
		\hline
		\textbf{Send Rate}
		& \textbf{Max Latency}
		& \textbf{Min Latency}
		& \textbf{Avg Latency}
		& \textbf{Throughput}
		\\ \hline
		
		912.3
		& 146.75
		& 57.47
		& 67.37
		& 483.7
		\\ \hline
	\end{tabularx}
\end{table*}

Table \ref{tab:createEHR_results} presents the average outcomes under observed CPU limitations after multiple benchmarks, highlighting instances where resource demands on both CPU and memory increase significantly. Table \ref{tab:caliper-tests} summarizes the results of random tests involving different worker configurations, varying transaction rates, and total transaction volumes, all aimed at querying EHR data via the chaincode.

\begin{table*}[htbp]
	\centering
	\caption{Selected Caliper Test Results for QueryEHR}
	\label{tab:caliper-tests}
	\begin{tabular}{ccc ccc c}
		\toprule
		\textbf{TxNum} & \textbf{Rate} & \textbf{Workers} & \textbf{Succ} & \textbf{Fail} & \textbf{Max Lat (s)} & \textbf{Thruput (TPS)} \\
		\midrule
		5000  & 2000 & 16 & 4840 & 152   & 43.69 & 109.9 \\
		15000 & 2000 & 32 & 2954 & 12022 & 70.93 & 194.9 \\
		20000 & 5000 &  8 & 7629 & 12371 & 91.90 & 207.3 \\
		20000 & 5000 & 32 & 7926 & 12074 & 86.72 & 211.5 \\
		50000 & 4000 & 16 &  949 & 49051 & 114.83 & 429.6 \\
		50000 & 5000 & 32 & 2844 & 47140 & 107.52 & 440.0 \\
		\bottomrule
	\end{tabular}
\end{table*}
 
Tables \ref{tab:createehr-perf},\ref{tab:createehr-performance} and \ref{tab:createehr-res} present the experimental results obtained from various test configurations aimed at creating new EHRs through the chaincode. The chaincode implementation involves multiple stages prior to appending the new EHR to the ledger.

\begin{table}[!htbp]
	\centering
	\caption{Performance Metrics of CreateEHR with Varying Workloads}
	\label{tab:createehr-perf}
	\begin{tabular}{lcccccccc}
		\toprule
		\textbf{Test} & \textbf{Workers} & \textbf{TxNum} & \textbf{TPS} & \textbf{Succ} & \textbf{Fail} & \textbf{MaxLat (s)} & \textbf{MinLat (s)} & \textbf{AvgLat (s)} \\
		\midrule
		\textbf{1} & 16 & 40000 & 5000 & 3015 & 36985 & 100.90 & 65.11 & 90.36 \\
		\textbf{2} & 32 & 70000 & 10000 & 514 & 69470 & 129.33 & 65.89 & 69.32 \\
		\textbf{3} & 16 & 50000 & 5000 & 13 & 49987 & 108.40 & 98.71 & 101.00 \\
		\bottomrule
	\end{tabular}
\end{table}

\begin{table}[!htbp]
	\centering
	\caption{Performance Metrics for \texttt{CreateEHR}}
	\label{tab:createehr-performance}
	\begin{tabular}{lccccccc}
		\toprule
		\textbf{Test} & \textbf{Succ} & \textbf{Fail} & \textbf{Send Rate} & \textbf{Max Lat.} & \textbf{Min Lat.} & \textbf{Avg Lat.} & \textbf{Throughput}\\
		\textbf{(Worker,TxNumber,TPS)} & & & \textbf{(TPS)} & \textbf{(s)} & \textbf{(s)} & \textbf{(s)} & \textbf{(TPS)} \\
		\midrule
		\textbf{1 (16,\,40000,\,5000)} & 3015 & 36985 & 1036.3 & 100.90 & 65.11 & 90.36 & 371.0 \\
		\textbf{2 (32,\,70000,\,10000)} & 514  & 69470 & 1060.8 & 129.33 & 65.89 & 69.32 & 528.0 \\
		\textbf{3 (16,\,50000,\,5000)} & 13   & 49987 & 1045.0 & 108.40 & 98.71 & 101.0 & 455.1 \\
		\bottomrule
	\end{tabular}
\end{table}

\begin{table}[!htbp]
	\centering
	\caption{Resource Usage for the Same Tests (Selected Chaincode Container)}
	\label{tab:createehr-res}
	\begin{tabular}{lcccccccc}
		\toprule
		\textbf{Test} & \textbf{CPU\% (max)} & \textbf{CPU\% (avg)} & \textbf{Mem (max) [MB]} & \textbf{Mem (avg) [MB]} & \textbf{In [MB]} & \textbf{Out [MB/KB]} \\
		\midrule
		\textbf{1} & 1.2125   & 0.14    & 20.6  & 13.7  & 9.01 & 2.54\,MB &  \\
		\textbf{2} & 0.2475   & 0.02125 & 12.7  & 7.68  & 1.47 & 446\,KB  \\
		\textbf{3} & 0.98125  & 0.18125 & 9.20  & 7.96  & 4.44 & 1.31\,MB \\
		\bottomrule
	\end{tabular}
\end{table}

Tables~\ref{tab:resource_usage_cpu_mem} and  \ref{tab:resource_usage_net_disk} present resource utilization. Notice that \texttt{/peer0.org1} exhibits higher CPU and memory usage compared to other nodes, which reflects heavier transaction loads. The contrasting network and disk I/O across containers also reveal how the workload is shared among the peers and the orderer.

\begin{table*}[!htbp]
	\centering
	\caption{Resource Usage (CPU and Memory)}
	\label{tab:resource_usage_cpu_mem}
	\begin{tabularx}{\textwidth}{|X|X|X|X|X|}
		\hline
		\textbf{Name}
		& \textbf{CPU\% (max)}
		& \textbf{CPU\% (avg)}
		& \textbf{Mem (max) [GB]}
		& \textbf{Mem (avg) [GB]}
		\\ \hline
		
		/dev-peer0.org1-cch7 
		& 9.3375 
		& 0.58125 
		& 0.372 
		& 0.213
		\\ \hline
		
		/dev-peer0.org2-cch7
		& 0.00125
		& 0
		& 0.0115
		& 0.0115
		\\ \hline
		
		/orderer
		& 5.08625
		& 0.59125
		& 0.174
		& 0.133
		\\ \hline
		
		/peer0.org2
		& 16.15875
		& 1.085
		& 0.128
		& 0.0951
		\\ \hline
		
		/peer0.org1
		& 82.94625
		& 19.525
		& 2.92
		& 2.26
		\\ \hline
		
	\end{tabularx}
\end{table*}

\begin{table*}[!htbp]
	\centering
	\caption{Resource Usage (Network and Disk)}
	\label{tab:resource_usage_net_disk}
	\begin{tabularx}{\textwidth}{|X|X|X|X|X|}
		\hline
		\textbf{Name}
		& \textbf{Net In [MB]}
		& \textbf{Net Out [MB]}
		& \textbf{Disk W [MB]}
		& \textbf{Disk R [MB]}
		\\ \hline
		
		/dev-peer0.org1-cch7
		& 35.0
		& 9.16
		& 0.00
		& 0.320
		\\ \hline
		
		/dev-peer0.org2-cch7
		& 0.00138
		& 0.00150
		& 0.00
		& 0.00
		\\ \hline
		
		/orderer
		& 10.3
		& 20.5
		& 23.8
		& 6.95
		\\ \hline
		
		/peer0.org2
		& 10.3
		& 0.199
		& 20.1
		& 0.410
		\\ \hline
		
		/peer0.org1
		& 78.3
		& 63.6
		& 20.1
		& 0.254
		\\ \hline
		
	\end{tabularx}
\end{table*}


\section{Conclusions and Future Research}
	\label{Conclusion}
	
	The proposed hChain 4.0 framework uses a permissioned blockchain for EHR management, ensuring strong security across all tiers. By integrating ABAC for fine-grained access control and PHE for secure computations on encrypted data, it further enhances privacy among diverse stakeholders. Multiple secured channels protect sensitive operations, while a tested throughput of 483 TPS supports real-time healthcare applications. Additionally, leveraging a permissioned design avoids public blockchain gas fees, making hChain 4.0 a cost-effective and practical choice for modern healthcare ecosystems.
	
	Future research will focus on developing an intuitive user interface to improve end-user engagement and expanding the system’s capacity to handle large, high-volume data. These enhancements will further solidify hChain 4.0 as a secure, private, and scalable solution for modern healthcare ecosystems.

\section*{Acknowledgment}
	This preprint is based on our conference paper hChain 4.0 \cite{hchain4}.

	\bibliographystyle{IEEEtran}
	\bibliography{Bibliography_hChain-4-0}  

\pagebreak

	\noindent
	\begin{minipage}[t]{0.14\textwidth}
		\vspace{0pt}
		\centering
		\includegraphics[width=0.9in,keepaspectratio]{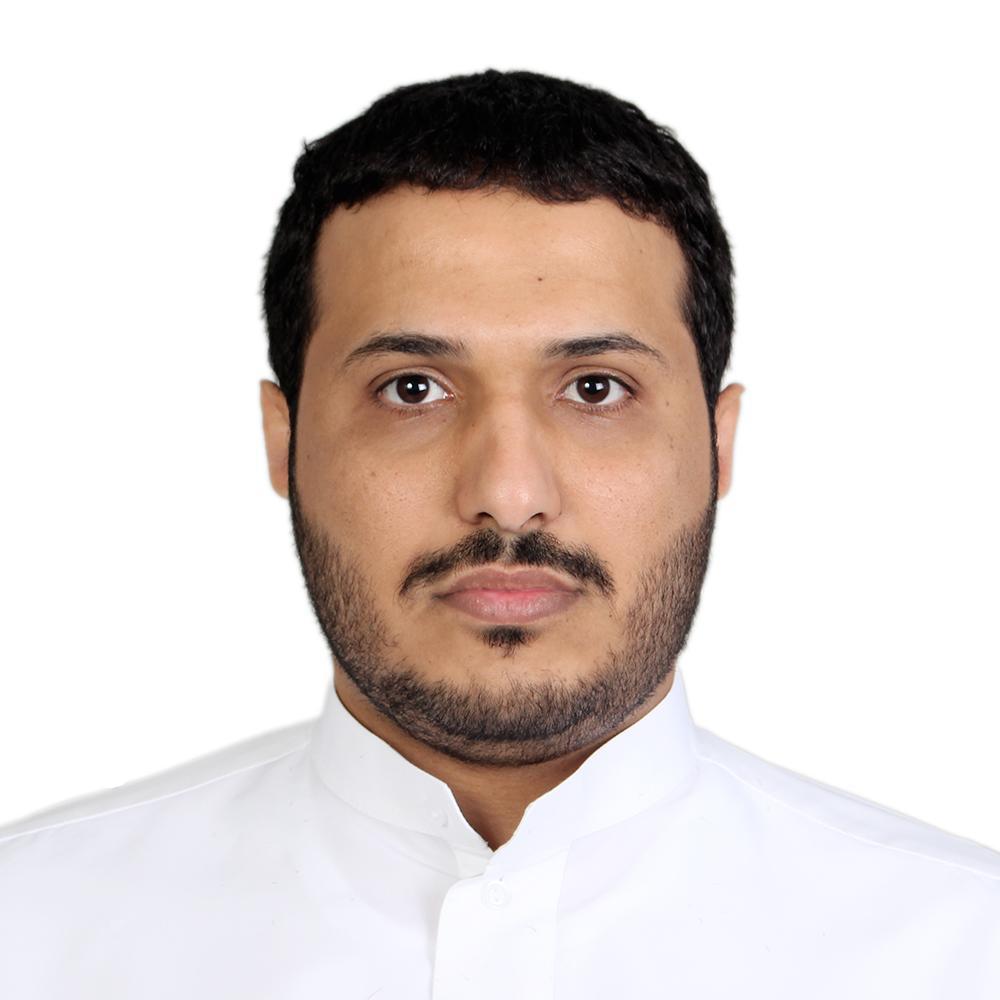}\\
	\end{minipage}%
	\hfill
	\begin{minipage}[t]{0.84\textwidth}
		\vspace{0pt}
		\footnotesize
		Musharraf Alruwaill is a Ph.D. Candidate in the Smart Electronic Systems Laboratory 
		(SESL) at the University of North Texas, under the mentorship of Dr. Saraju Mohanty. 
		He received his Bachelor’s degree from Jouf University and later earned his Master 
		of Science in Computer Science at the University of New Haven. With six publications 
		to his credit, his research interests focus primarily on developing and implementing 
		advanced technologies for smart city applications.
	\end{minipage}

\vspace{1.0cm}
	\noindent
	\begin{minipage}[t]{0.14\textwidth}
		\vspace{0pt}
		\centering
		\includegraphics[width=0.9in,keepaspectratio]{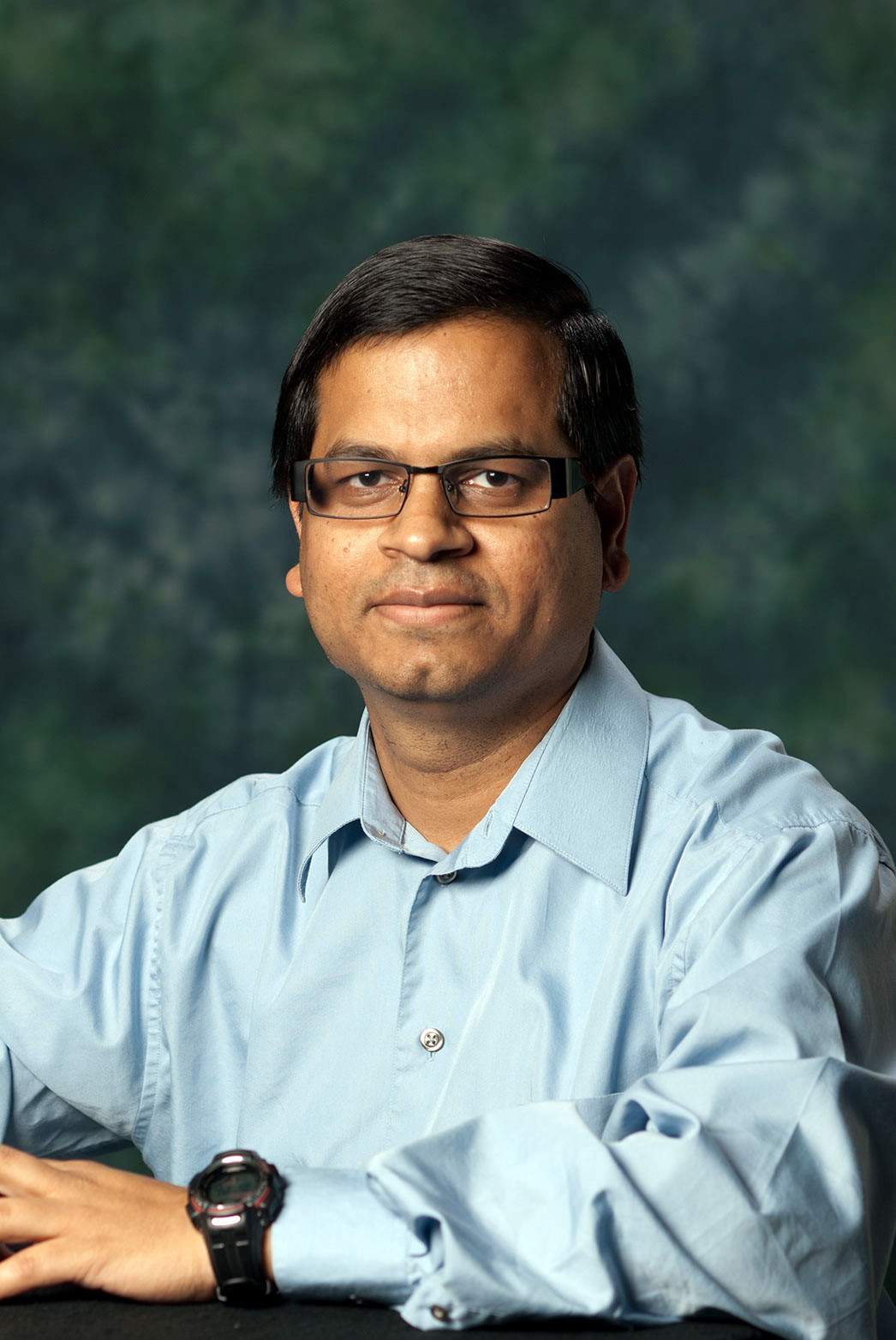}\\
	\end{minipage}%
	\hfill
	\begin{minipage}[t]{0.84\textwidth}
		\vspace{0pt}
		\footnotesize
		Saraju P. Mohanty (Senior Member, IEEE) received the bachelor’s degree (Honors) in electrical engineering 
		from the Orissa University of Agriculture and Technology, Bhubaneswar, in 1995, the master’s degree in Systems 
		Science and Automation from the Indian Institute of Science, Bengaluru, in 1999, and the Ph.D. degree in 
		Computer Science and Engineering from the University of South Florida, Tampa, in 2003. He is a Professor with 
		the University of North Texas. His research is in ``Smart Electronic Systems,'' which has been funded by the 
		National Science Foundation, Semiconductor Research Corporation, the U.S. Air Force, NIDILRR, IUSSTF, and 
		Mission Innovation. He has authored 550 research articles, 5 books, and 10 granted or pending patents. His 
		Google Scholar h-index is 62 and i10-index is 298, with 16,000 citations. He is regarded as a visionary 
		researcher on Smart Cities technology, focusing on security, energy-aware design, and AI/ML integration.
		
		He introduced the Secure Digital Camera (SDC) in 2004 with built-in security features using 
		Hardware Assisted Security (HAS) or Security by Design (SbD) principles. He is widely credited as the 
		designer of the first digital watermarking chip in 2004 and the first low-power digital watermarking chip 
		in 2006. Among his accolades are 21 best paper awards, a Fulbright Specialist Award in 2021, the IEEE 
		Consumer Electronics Society Outstanding Service Award in 2020, the IEEE-CS-TCVLSI Distinguished 
		Leadership Award in 2018, and the PROSE Award for Best Textbook in Physical Sciences and Mathematics 
		in 2016. He has delivered 31 keynotes and served on 15 panels at various international conferences.
		
		He has been on the editorial boards of several peer-reviewed international transactions/journals, 
		including \textit{IEEE Transactions on Big Data}, \textit{IEEE Transactions on Computer-Aided Design of 
			Integrated Circuits and Systems}, \textit{IEEE Transactions on Consumer Electronics}, and the 
		\textit{ACM Journal on Emerging Technologies in Computing Systems}. He served as Editor-in-Chief of the 
		\textit{IEEE Consumer Electronics Magazine} from 2016 to 2021, chaired the Technical Committee on Very 
		Large Scale Integration for the IEEE Computer Society from 2014 to 2018, and was on the Board of Governors 
		of the IEEE Consumer Electronics Society from 2019 to 2021. Presently, he serves on the steering, organizing, 
		and program committees of multiple international conferences, including \textit{IEEE International Symposium 
			on Smart Electronic Systems} (IEEE-iSES), \textit{IEEE-CS Symposium on VLSI} (ISVLSI), and \textit{OITS 
			International Conference on Information Technology} (OCIT). Over his career, he has supervised 3 postdoctoral 
		researchers, 18 Ph.D. dissertations, 28 M.S. theses, and 41 undergraduate projects.
	\end{minipage}
	
\vspace{1.0cm}
	\noindent
	\begin{minipage}[t]{0.14\textwidth}
		\vspace{0pt}
		\centering
		\includegraphics[width=0.9in,keepaspectratio]{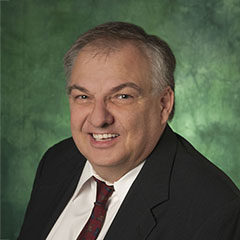}\\
	\end{minipage}%
	\hfill
	\begin{minipage}[t]{0.84\textwidth}
		\vspace{0pt}
		\footnotesize
		Elias Kougianos received a BSEE degree from the University of Patras, Greece in 1985 
		and an MSEE in 1987, an MS in Physics in 1988, and a Ph.D. in Electrical Engineering 
		in 1997, all from Louisiana State University. From 1988 to 1998, he was with Texas 
		Instruments, Inc. in Houston and Dallas, TX. In 1998, he joined Avant! Corp. (now 
		Synopsys) in Phoenix, AZ, as a Senior Applications Engineer, and in 2000 he moved 
		to Cadence Design Systems, Inc. in Dallas, TX, serving as a Senior Architect for 
		Analog/Mixed-Signal Custom IC design.
		
		Since 2004, he has been with the University of North Texas (UNT), where he is 
		currently a Professor in the Department of Electrical Engineering. His research 
		interests include Analog/Mixed-Signal/RF IC design and simulation, as well as 
		the development of VLSI architectures for multimedia applications. He has authored 
		more than 200 peer-reviewed journal and conference publications.
	\end{minipage}

\end{document}